\documentclass[a4paper,prl,twocolumn,showpacs,10pt,aps,reprint]{revtex4-1}
\usepackage{amsmath,amssymb,hyperref}
\usepackage{graphicx}
\usepackage{datetime}
\usepackage{braket}
\usepackage[usenames,dvipsnames]{color}

\definecolor{MyGreen}{rgb}{0,0.6,0}

\newcommand{\figref}[1]{Fig.~\ref{#1}}
\newcommand{\eqnref}[1]{Eqn.~(\ref{#1})}

\newcommand{\micron}{\ensuremath{\mu{\rm m}}}
\newcommand{\drr}{\ensuremath{{\rm d}^2{\bf r}}}
\newcommand{\dpp}{\ensuremath{{\rm d}^2{\bf p}}}
\newcommand{\rp}{\ensuremath{{\bf r,p}}}
\newcommand{\dirac}[1]{\ensuremath{\delta\left[ {#1}\right]}}
\newcommand{\expo}[1]{{\rm e}^{#1}}
\newcommand{\fboltz}[1]{\expo{\left[{-\frac{{#1}}{k_B T}}\right]}}

\begin{document}

\title{Phase-space views into dye-microcavity thermalised and condensed photons}

\author{Jakov Marelic}
\author{Benjamin T. Walker}
\author{Robert A. Nyman}\email[Correspondence to
]{r.nyman@imperial.ac.uk} \affiliation{Quantum Optics and Laser Science group,
Blackett Laboratory, Imperial College London, Prince Consort Road, SW7
2BW, United Kingdom}


\begin{abstract}
	We have observed momentum- and position-resolved spectra and images of the photoluminescence from thermalised and condensed dye-microcavity photons. The spectra yield the dispersion relation and the potential energy landscape for the photons. From this dispersion relation, we find that the effective mass is that of a free photon not a polariton. We place an upper bound on the dimensionless two-dimensional interaction strength of $\tilde{g}\lesssim 10^{-3}$, which is compatible with existing estimates. Both photon-photon and photon-molecule interactions are weak. The temperature is found to be independent of momentum, but dependent on pump spot size, indicating that the system is ergodic but not perfectly at thermal equilibrium. Condensation always happens first in the mode with lowest potential and lowest kinetic energy, although at very high pump powers multimode condensation occurs into other modes.
\end{abstract}

\maketitle

%
%


The concept of a phase-space distribution played an important role in the development of statistical mechanics before quantum mechanics divided the universe into discrete states~\cite{Gibbs}\footnote{We would have cited Ludwig Boltzmann here but our German language skills are insufficient to find which of his articles is most appropriate.}. The Boltzmann equation~\cite{Huang} describes the evolution of the phase-space distribution for a system made of particles which interact among themselves, through a term known as the collision integral. For indistinguishable quantum particles, the collision integral can be modified and the Boltzmann equation is still valid\cite{Wu97}. There exist time-invariant phase-space distributions which are functions only of a phase-space local energy (called ``ergodic'') for which entropy is maximised for some fixed control parameters (called ``thermal equilibrium'').  At thermal equilibrium, the statistical distribution can be shown on purely information-theoretic grounds to be independent of the physical phenomena by which equilibrium is obtained~\cite{Jaynes57}.

True thermal equilibrium takes an infinite time to be achieved. In dynamical thermalising systems with finite lifetime, thermal equilibrium cannot therefore occur except approximately. Two important examples of dynamical-equilibrium quantum systems are trapped atomic gases and microcavity polaritons. Ultracold atoms can be modelled using the Boltzmann equation~\cite{Luiten96}, with the collision integral directly evaluated, for example using Monte Carlo molecular-trajectory methods~\cite{Wu96}. Loss is considered as a perturbation, either a low-probability event or a high-energy truncation of a near-equilibrium distribution. Microcavity polaritons, even when non-resonantly driven, can be driven far from equilibrium, and may show very non-thermal steady states~\cite{Kasprzak08}.

Photons in a dye-filled microcavity~\cite{Klaers10a, Klaers10b} are unusual in that the thermal equilibrium process is not among particles, but between particles and a thermal bath~\cite{deLeeuw13, Chiocchetta14} via the vibrational relaxation of the dye molecules. Interparticle interactions are very weak so the collision integral is effectively zero~\cite{Nyman14, VanDerWurff14}, yet still bosonic exchange statistics lead to condensation~\cite{Klaers10b, Snoke13}. A classical rate equation for the populations of states can be derived from a quantum master equation~\cite{Kirton13, Kirton15, Keeling16}, but the Boltzmann equation itself does not directly apply, because the collision integral would be zero. Non-equilibrium distributions have been observed with time-dependent~\cite{Schmitt15} or steady-state~\cite{Marelic15, Marelic16} pumping. Approximate thermal equilibrium is obtained if the scattering rate for photons from dye molecules is faster than the cavity loss rate, and under those circumstances standard thermodynamic properties can be observed, such as the lambda transition associated with Bose-Einstein condensation~\cite{Damm16}. 

One can think of the photoluminescence emitted from the microcavity as representing the phase-space distribution of the intra-cavity photons. A position-space image is the phase-space distribution marginalised over all momenta. Similarly the observed spectrum can be thought of as the phase-space distribution integrated over the phase space wherever the local-energy function is matched. The spectrum is not always sufficient to determine whether states are occupied thermally, since the density of states may not be known. First- and second-order correlations are simply related in thermal equilibrium, and dye-microcavity photons have been seen to be compatible with thermal equilibrium~\cite{Schmitt16}. 

We begin this manuscript with a discussion of phase-space distributions for non-interacting thermal-equilibrium distributions, and how they can be observed. We then describe our experimental setup for dye-thermalised microcavity photons, including the optics required for various phase-space views. We present momentum-resolved spectra both below and above threshold pump power for condensation. From the observed dispersion relation we infer upper bounds for the strength of photon-molecule and photon-photon interactions. We also show momentum-space images, noting that condensation happens in the lowest-momentum, lowest-energy mode. Turning to position-resolved spectra, we map the potential energy landscape.

\section{Views into phase-space at thermal equilibrium}

The phase-space distribution, the population as a function of momentum for all positions,  of a non-interacting gas contains all of the relevant information about the gas, when combined with the phase-space local energy. Non-equilibrium, or non-ergodic systems can be fully described via the phase-space distribution. We shall write a general phase-space distribution as $f(\rp)$, as function of position ${\bf r}$ and momentum ${\bf p}$. The microcavity used for thermalising photons permits two free dimensions, but the following discussion can easily generalised to other dimensionalities. 

One simple view into phase space is a position-space image $n({\bf r})$ which is simply $f(\rp)$ marginalised over all momenta:
\begin{align}
  n({\bf r}) = \int \dpp \,f(\rp)
\end{align}
Similarly the observed spectrum $n(E)$ can be thought of as the phase-space distribution integrated over the phase-space wherever the local-energy function $\epsilon(\rp)$ is matched~\footnote{Here we are making an approximation that the phase-space distribution varies slowly compared to discreteness of states, i.e. Planck's constant is small.}: 
\begin{align}
  n(E) = \iint \drr\,\dpp\,f(\rp)\dirac{E- \epsilon(\rp)}
\end{align}
The position-resolved spectrum $n({\bf r},E)$ is an alternative view into phase space:
\begin{align}
  n({\bf r},E) = \int \dpp\,f(\rp)\dirac{E- \epsilon(\rp)}
\end{align}
The momentum-resolved spectrum, $n({\bf p},E)$ is determined similarly, marginalising over positions.

Thermal equilibrium distributions are determined by the phase-space local energy only.  For light in a near-planar microcavity the local potential energy $V({\bf r})$ is determined by the shape of the mirrors that form the cavity~\cite{Lugiato87, Chiao99, Chiao00}. 
The kinetic energy $Q({\bf p})$ is momentum-space local. For non-interacting particles of mass $m$, \mbox{$Q({\bf p}) = \frac{p^2}{2 m}$}. The total energy is then \mbox{$\epsilon(\rp) = Q({\bf p}) + V({\bf r})$}.

For example the Boltzmann distribution of occupancies of states of energy $E$ is $f(\rp) = \frac{1}{Z}\fboltz{\epsilon(\rp)}$, where $\iint \drr\,\dpp\, f(\rp) = 1$ defines the partition function $Z$, and $T$ is the temperature. It is straightforward to show that, since the energy is a quadratic, radially symmetric function of the magnitude of momentum in two dimensions for free massive particles:
\begin{align}
	n({\bf r}, E) = 
	\begin{cases}
		\frac{2\pi m}{Z} \fboltz{E} &\text{, if } E>V({\bf r}) \\
				0 &\text{ otherwise}
	\end{cases}
	\label{eqn:frE}
\end{align}
The position-resolved spectrum then gives information about both the thermal distribution, through the temperature, and the potential energy landscape, through a position-dependent cutoff in the spectrum, below which no light is detectable. 

The momentum-resolved spectrum is defined similarly:
$
  n({\bf p},E) = \int \drr\,f(\rp)\dirac{E- \epsilon(\rp)}.
$
In the case of a quadratic radially symmetric potential (the isotropic harmonic oscillator, typically the case for dye-microcavity thermalised photons) with \mbox{$V({\bf r}) = \frac{1}{2}m\Omega^2 r^2$} for trap angular frequency $\Omega$, the momentum-resolved spectrum looks very similar to the position-resolved spectrum:
\begin{align}
	n({\bf p}, E) = 
	\begin{cases}
		\frac{2\pi}{Z m\Omega^2} \fboltz{E} &\text{, if } E>\frac{p^2}{2m} \\
				0 &\text{ otherwise}
	\end{cases}
	\label{eqn: momentum spec}
\end{align}
and the dispersion relation can be detected through the momentum-dependent cutoff. For distributions other than the Boltzmann, the exponential energy factor is replaced by the appropriate function, e.g $\left[{\rm e}^{(E-\mu)/kT} - 1 \right]^{-1}$ for a Bose-Einstein distribution with chemical potential $\mu$. Phase space must be discretised carefully in units of the Planck constant.

These principles apply to photons in a dye-filled microcavity. In the limit of weak photon-photon interactions, as well as paraxial approximation, the photons have an effective mass $m_{ph}$ given by the cavity length: $m_{ph} = h\,n^2 / (c \lambda_0)$ where $\lambda_0$ is the wavelength (in free space) associated with light of the energy of the lowest available cavity mode~\cite{Nyman14}, which we will call the cutoff wavelength. Here $n$ is the refractive index of the medium, $c$ the speed of light in free space and $h$ Planck's constant. The potential energy landscape is defined by the shape of the cavity mirrors. It is straightforward to show that for a plano-spherical cavity $V({\bf r}) =\left(\frac{h c}{\lambda_0}\right) \frac{1}{2 L_0 R} r^2$ with $L_0$ being the length of the cavity at its greatest, and $R$ the mirror radius of curvature, so that a harmonic trap is formed of characteristic angular frequency $\Omega = (c/n) / \sqrt{L_0 R}$. 
In a planar-planar microcavity, the potential energy is a constant and $f({\bf p}, E) = \frac{1}{Z}\fboltz{E}\dirac{E-Q({\bf p})}$, i.e. the photoluminescence is concentrated on a line which follows the dispersion relation, weighted by the Boltzmann factor. 

Dispersion relations have been measured in microcavities, for example showing Bose-Einstein condensation of exciton-polaritons~\cite{Kasprzak06} and organic polaritons~\cite{Daskalakis14, Plumhof14}, the Bogoliubov dispersion relation for interacting Bose condensates~\cite{Utsunomiya08} as well as the deviation from the simple massive-particle dispersion induced by strong coupling between light and matter~\cite{Houdre94, Lidzey99}. It is worth noting that in ultracold atoms~\cite{Stenger99, Steinhauer03} and superfluid helium~\cite{Donnelly81}, an equivalent quantity is often referred to as the structure factor.

\section{Experimental setup}

\begin{figure*}[htb]
	\centering
	\includegraphics[width=1.3\columnwidth]{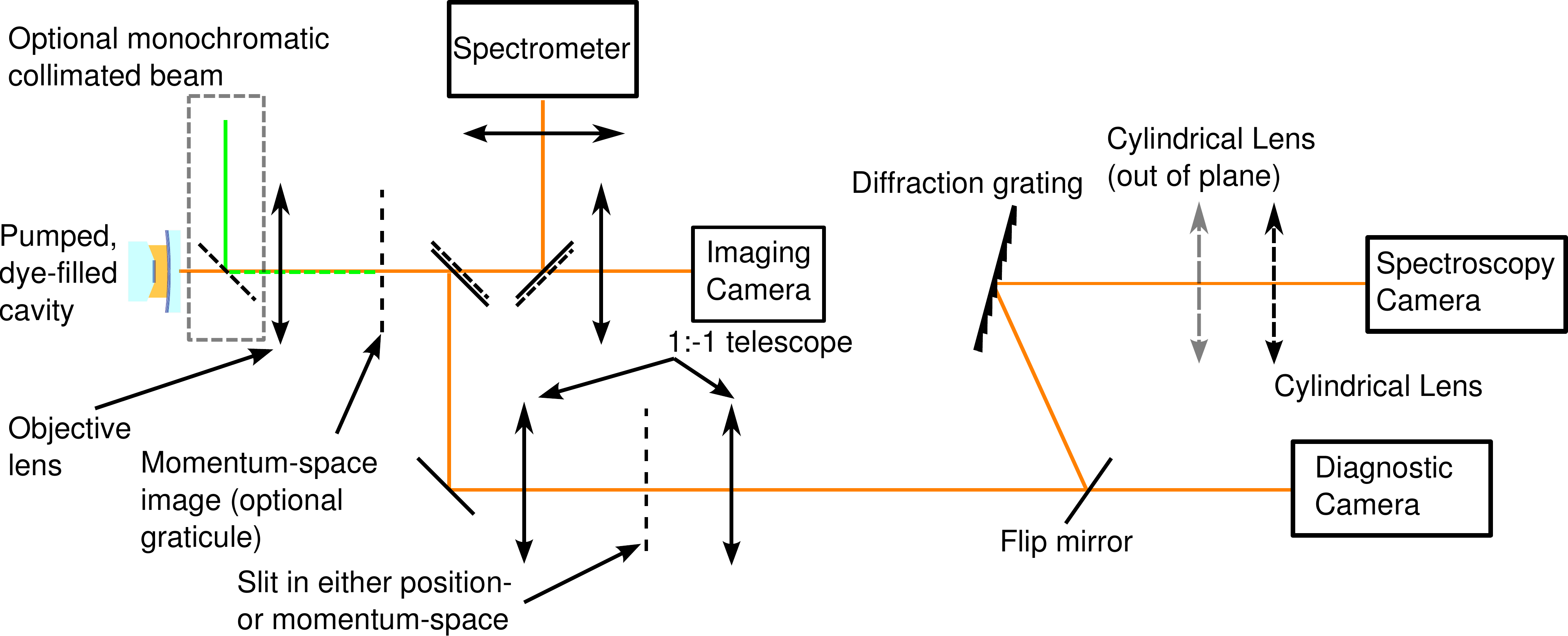}
	\includegraphics[width=0.7\columnwidth]{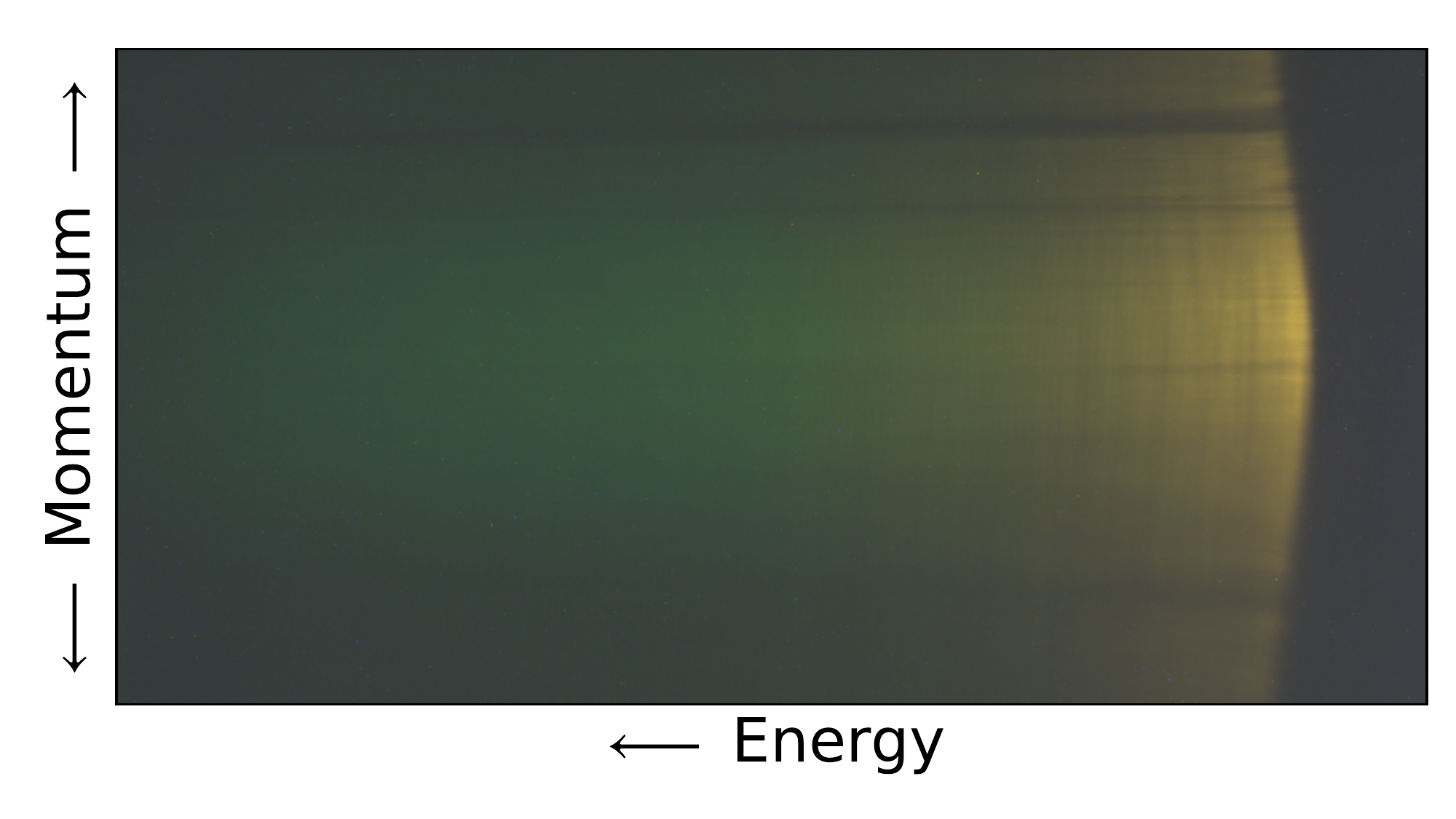}
	\caption{Left: a schematic of the optics for momentum- and position-resolving spectroscopy. Double solid-dashed lines are beamsplitters. Lenses are marked with double-headed arrows. Right: an example momentum-resolved spectrum. The only post-processing performed on the colour image is taking the square-root for display purposes. Calibration of the axes is described later in this manuscript. Dark lines are dirt stuck on the slit in momentum space.}
	\label{fig:setup}
\end{figure*}

The core of our experimental apparatus consists of the same optically-pumped, dye-filled microcavity described in our previous articles~\cite{Marelic15, Marelic16}: see \figref{fig:setup}~\footnote{The planar mirror used is the same as before, but the spherical mirror has a different specification for the reflectivity coating. It now lets through 30~ppm of light, marginally limiting the cavity finesse but giving increased signal strength.}. The light is chopped into 500~ns pulses using an acousto-optical modulator (AOM), with the pulse repetition rate lower for higher pump powers, ensuring that few dye molecules are shelved in the unwanted triplet state. The cavity thermalised light is in one longitudinal mode, typically the 8th mode, so that the cavity is 4 wavelengths of light long. The photoluminescence is imaged to infinity by a objective lens, and the light split on a beamsplitter (shown as a double solid-dashed line), with half sent to be shared among an imaging camera and a commercial spectrometer. The rest of the light goes to the resolving spectroscopy optics, which can be configured in multiple ways for measurement and calibration. The main resolving spectroscopy components are: a telescope which creates images for position- or momentum-spatial filtering; a diffraction grating to separate wavelengths of light; and cylindrical optics which focus the light onto the spectroscopy camera. Intra-cavity in-plane momentum corresponds to angle of emission, implying that the momentum image is the Fourier transform of the position-space.

The light can also be sent directly to a diagnostic camera, which is initially focussed on position-space features of the microcavity, usually one edge of the 1~mm diameter planar mirror. A 1:1 telescope is then inserted into the optical path, and focussed such that the image on the diagnostic camera is re-created. The telescope uses a pair of 50~mm focal length achromatic doublets (the same as the objective) to reduce aberrations, both chromatic and off-axis.

A slit is placed in the telescope, in one of two positions, either the position- or momentum-space plane. The 20~\micron\ wide slit is vertical, reducing a 2D image to one dimension. To find the position-space plane, we send photoluminescence through the slit and move the slit until its image is in focus on the diagnostic camera. We use collimated light, inserted just before the objective as a single-momentum light source for placing the slit in momentum-space. To focus the space (vertical) axis of the spectroscopy camera we rotate the slit to horizontal, and focus transmitted photoluminescence to a thin horizontal line, then rotate the slit back to vertical. 


We focus the energy (horizontal) axis after the vertical axis, and use a photon Bose-Einstein condensate, which is known to be narrow-band just above threshold pump power~\cite{Schmitt16} as a single-energy reference light source. Calibration of energy uses the commercial spectrometer as a reference to generate two known pixel-energy pairs. We then extend the cavity to about 100 wavelengths of light, and use the many longitudinal modes as a comb of frequency references, to which we fit a cubic calibration of energy as a function of pixel number. The calibration is nearly linear, with the gradient varying by no more than 10\% across the image. The wavelength resolution is limited by vibrations of the microcavity, and can be as large as 0.5~nm. For the very shortest cavities (not used in this article) the mirrors bump together reducing vibrations, leaving the resolution, measured using a photon condensate, as good as 0.05~nm, which is equivalent to just 2 pixels on camera.

\section{Momentum space}

\subsection{Calibration of momentum}


The back focal plane of the objective lens is a momentum-space image plane. The correspondence between displacement from the optic axis in the back focal plane $y_{bfp}$  and vertical intracavity momentum $p_y$ is:
\begin{align}
  p_y = y_{bfp} \times \frac{h}{\lambda_0 f'}
  \label{eqn:momentum calibration}
\end{align}
where $f'$ is the focal length of the objective lens. In deriving this relation, we are required to make use of the Minkowski formula for canonical momentum ($h/\lambda$) in a medium, not the Abraham formula for kinetic momentum~\cite{Barnett10}. Near the back focal plane of the objective, we can insert a calibrated graticule of black lines printed on acetate. The known size of the graticule features provides the calibration, even though the magnification of the imaging system from back focal plane to camera is not \textit{a priori} known. The measured magnification is very insensitive to the exact position of the graticule, so we place it only approximately in the back focal plane of the objective.

\begin{figure}[htb]
	\centering
	\includegraphics[width=0.45\columnwidth]{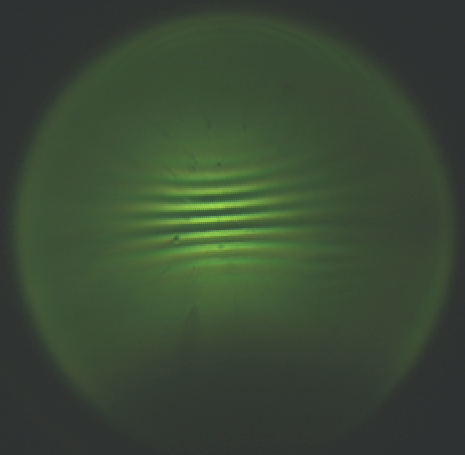}
	\includegraphics[width=0.53\columnwidth]{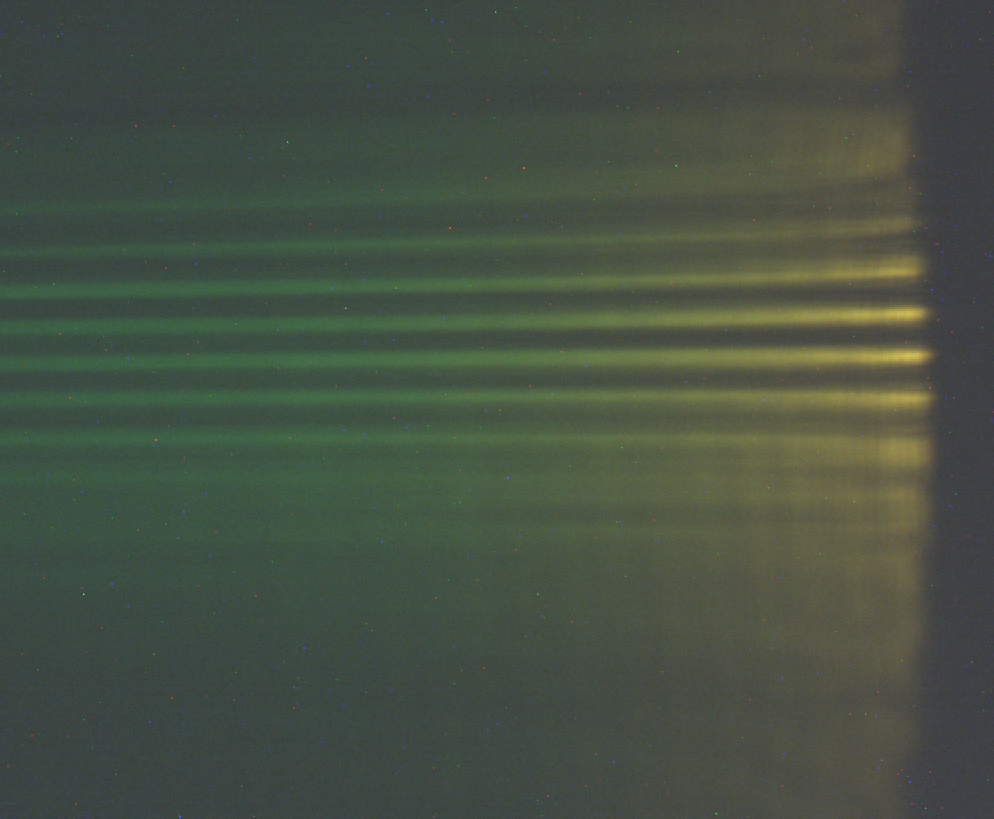}
	\caption{Momentum calibration using a graticule with 800~\micron\ pitch in the back focal of the imaging objective. Each line corresponds to a difference of in-plane $1.8\times10^{-29}$~N\,s. Left: a momentum space image. Right: momentum-resolved spectroscopy, with the horizontal axis being energy increasing to the left.}
	\label{fig:calibration momentum}
\end{figure}

In \figref{fig:calibration momentum} we show, with the graticule in place, an image in 2D momentum space (left) and the momentum-resolved spectrum of the cavity photoluminescence (right). The images are real colour, but the square-root of the raw value for display. The graticule had a pitch of 800~\micron, the cutoff wavelength we use is $\lambda_0 \simeq 588$~nm and the objective has a nominal focal length of 50~mm, so from one line to the next corresponds to a difference of $1.8\times10^{-29}$~N\,s. The momentum resolution in the centre is approximately 5 times smaller than the spacing between lines: $4\times10^{-30}$~N\,s.

Both images show distortions. The alignment of the light path, especially through the telescope, is centred to avoid astigmatism. The central undistorted portion shows a linear calibration between camera position and momentum up to about $4\times 10^{-29}$~N\,s. For larger momenta, some pincushion distortion which is typical of spherical aberration is noticeable. The momentum-resolved spectrum shows a change in magnification with energy relative to the cutoff energy, which persists when the cutoff energy is changed by changing the cavity length. This distortion means that our calibration is uncertain to within about 30\%, and it probably comes from position-space off-axis aberrations. 

\subsection{Momentum-resolved spectra: $n({\bf p}, E)$}

\begin{figure}
	\centering
	\includegraphics[width=0.98\columnwidth]{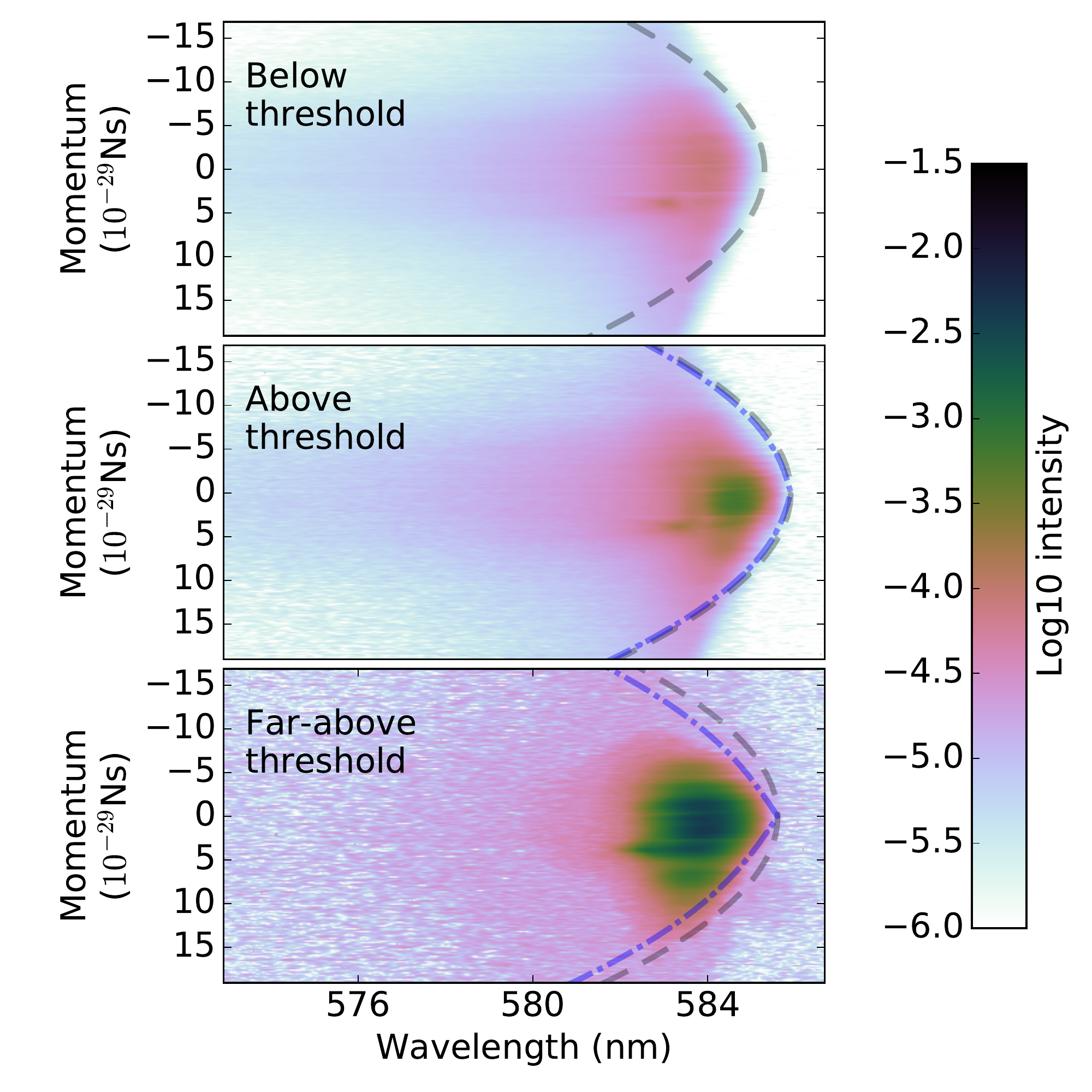}
	\caption{Momentum-resolved spectra $n({\bf p}, E)$ of photoluminescence below threshold (top) and BEC just above (middle) and far above threshold (bottom). Pump spot size was $70\,\micron$, optimised to produce a thermal distribution of photons at room temperature. The grey dashed lines are dispersion relations for particles of mass $7.8\times 10^{-36}$~kg. The blue dot-dashed lines on the lower two panels are the Bogoliubov dispersion relations assuming a dimensionless two-dimensional interaction parameter $\tilde{g} = 10^{-4}$ (see text for more details).}
	\label{fig:thermal momentum resolved}
\end{figure}

In \figref{fig:thermal momentum resolved} we show calibrated, false-colour momentum-resolved spectra, $n({\bf p}, E)$. Background light as well as camera hot pixels are eliminated by subtracting a background image, taken under the same conditions but with the pump light shut off by the AOM.

The expected dispersion relation for microcavity photons can be thought of as the requirement for light with in-plane momentum to satisfy the boundary conditions at the mirror surfaces. Alternatively, in the Schr\"odinger equation which describes the motion of the photons, the effective mass comes from the coefficient of the spatial derivative term, i.e. the kinetic energy. The grey parabola on the top panel of \figref{fig:thermal momentum resolved} represents the dispersion relation for a free particle of mass $m_{ph} =7.8\times 10^{-36}$~kg, using $n=1.44$ for the refractive index of the intracavity medium (mostly ethylene glycol).


At higher pump powers, Bose-Einstein condensation occurs. We see in \figref{fig:thermal momentum resolved} (lower panels) that the macroscopic occupation occurs in the lowest energy state, for low momentum.

\begin{figure}[htb]
	\centering
	\includegraphics[width=0.9\columnwidth]{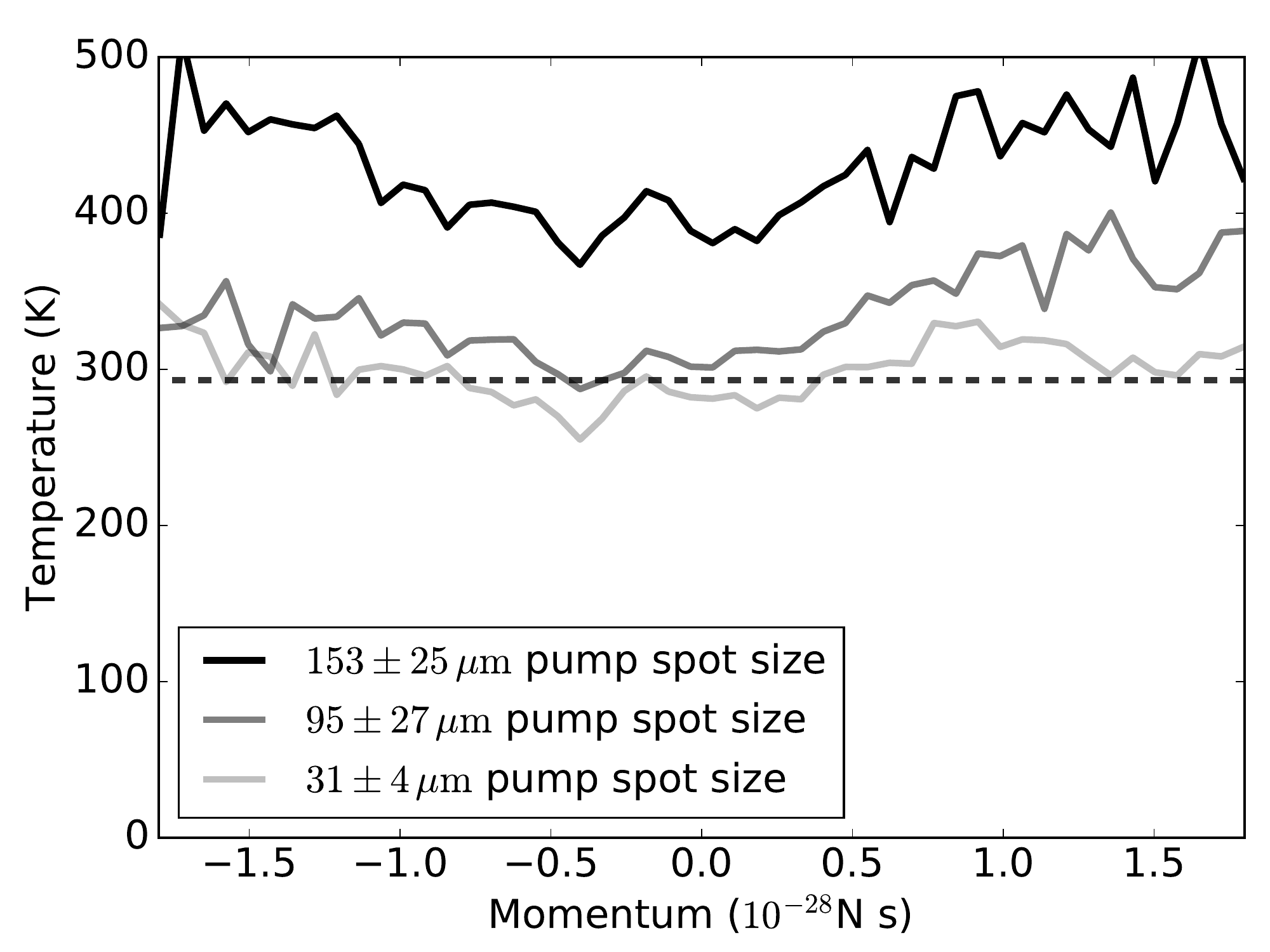}
	\caption{Temperature vs. momentum for several pump spot sizes, derived from momentum-resolved spectra. Thermalisation is not complete, and but larger spot sizes lead to higher temperatures. For comparison, the thermal-equilibrium cloud size is 70~\micron. Temperature is independent of momentum, implying that populations depend only on energy so ergodicity is respected.
	}
	\label{fig:temperature vs momentum many pump sizes}
\end{figure}

Below threshold, we have tested the validity of \eqnref{eqn: momentum spec}. For each momentum, we fitted the spectrum with an exponential decay with a cutoff. The cutoff is directly related to the dispersion relation. From \figref{fig:thermal momentum resolved} (top panel), it is clear that the dispersion relation follows that of a massive particle for small momenta. For large momenta (large emission angles), the pincushion distortion of our imaging system prevents us from drawing any conclusion. The exponential decay constant is the inverse temperature, as plotted in \figref{fig:temperature vs momentum many pump sizes}. The temperature is largely independent of momentum, compatible with \eqnref{eqn: momentum spec}, but the temperature does depend on the pump spot size. It is known that equilibration with room temperature is not complete, but limited by spontaneous emission from the dye molecules out of the cavity~\cite{Keeling16}. A similar conclusion can be drawn from the observation that the position-space size of the photon cloud depends on the pump spot size~\cite{Marelic15}. The implication is that even when thermal equilibrium is broken, ergodicity is still respected, i.e. a distribution in terms of energy alone is enough to describe the system.

\subsubsection{Effective mass}

If the photons are strongly coupled to molecules of mass $m_{molecule}$, they form polaritons, whose effective masses can be written in terms of the Hopfield coefficients~\cite{Deng10}: 
\mbox{
$|X_\pm|^2 = \frac{1}{2} \left\{ 1 \pm  \left[\frac{\Delta }{\sqrt{\Delta^2 + 4g_0^2}}\right] \right\}$
}
where $\Delta$ is the cavity detuning from the molecular resonance, known as the zero-photon line (ZPL) and $g_0$ the light-molecule coupling strength. The effective masses of the two kinds of polaritons (usually known as ``upper'' and ``lower'' polaritons) are
\begin{align}
  \frac{1}{m_\mp} = \frac{|X_\pm|^2}{m_{molecule}} + \frac{|X_\mp|^2}{m_{ph}}.
\end{align}

Within the range of validity of the calibration of momentum, the below-threshold dispersion relation matches the kinetic energy of a microcavity photon of the expected mass, $m_{ph}$. We have measured this effective mass for cutoff wavelengths between 520 and 610~nm. Data were taken within 1~nm of the ZPL at 545~nm. We found the effective mass to be compatible with the bare photon mass to within the variation of wavelength calibration: about 30\% across the camera. From this, we conclude that the coupling strength $g_0 < 1$~THz, and that the particles involved are really un-dressed cavity photons and not polaritons.

From the known spontaneous emission lifetime of 4~ns and the volume of the smallest cavity mode, $4\times 10^{-17}$~m$^3$, we expect the coupling of a single molecule to the cavity mode to be of order 1~GHz~\cite{Loudon}. There are at least $10^7$ molecules within a cavity mode. It seems that there is no collective enhancement of the light-matter coupling, as expected here since the same mechanism which helps photons thermalise also de-phases the electronic states of the molecules.


\subsubsection{Photon-photon Interactions}

Interactions are expected to cause the dispersion relation to deviate from parabolic, following the Bogoliubov dispersion. In a local-density approximation, that dispersion relation is:
\begin{align}
  Q({\bf p}) = \frac{p^2}{2m_{ph}} \left(1 + \frac{4\hbar^2\tilde{g}\rho_{ph}}{p^2} \right)^{1/2}\label{eqn:Bogoliubov}
\end{align}
where $\tilde{g}$ is the dimensionless two-dimensional interaction parameter and $\rho_{ph}$ the number density of photons in the centre of the condensate.

The number of photons in the condensate $N_C$ was estimated from the commercial spectrometer data (not shown), being about 12\% and 70\% of the non-condensate photon number for middle (just above threshold) and bottom (far-above threshold) panels respectively. We assume that the total number of non-condensate photons is given by the equilibrium saturation value \mbox{$N_{th} = \frac{pi^2}{6} \left(\frac{k_B T}{\hbar\Omega}\right)^2 \simeq 27\,000$}, for mode number $q=8$ and mirror radius of curvature 250~mm. The central two-dimensional density is given by the condensate number and harmonic oscillator length $a_{ho} = \sqrt{\hbar / m_{ph}\Omega}\simeq6\,\micron$ here: $\rho_{ph} = N_C / 2\pi a_{ho}^2$. 

In \figref{fig:thermal momentum resolved} we have overlaid the Bogoliubov dispersion relation (blue dot-dashed lines) onto the momentum-resolved spectra, for $\tilde{g} = 10^{-4}$. Larger values of $\tilde{g}$ are not compatible with the observed data in the bottom panel. Given that the local-density approximation is not valid for such weak interactions, that the condensate is probably multimode, and that imperfections in the cavity mirrors distort the condensate, we conclude a more conservative, loose upper bound of $\tilde{g}\lesssim 10^{-3}$. This value is consistent with previously published estimates based on experimental data which are in the range $7\times 10^{-4}$~\cite{Klaers10b} to $5\times 10^{-7}$~\cite{VanDerWurff14}. 
Interactions might alternatively be inferred by observing frequency shifts in anisotropic potentials~\cite{deLeeuw16, Vyas14}.

\subsection{Momentum-space images: $n({\bf p})$}

\begin{figure}[htb]
	\centering
	\includegraphics[width=0.8\columnwidth]{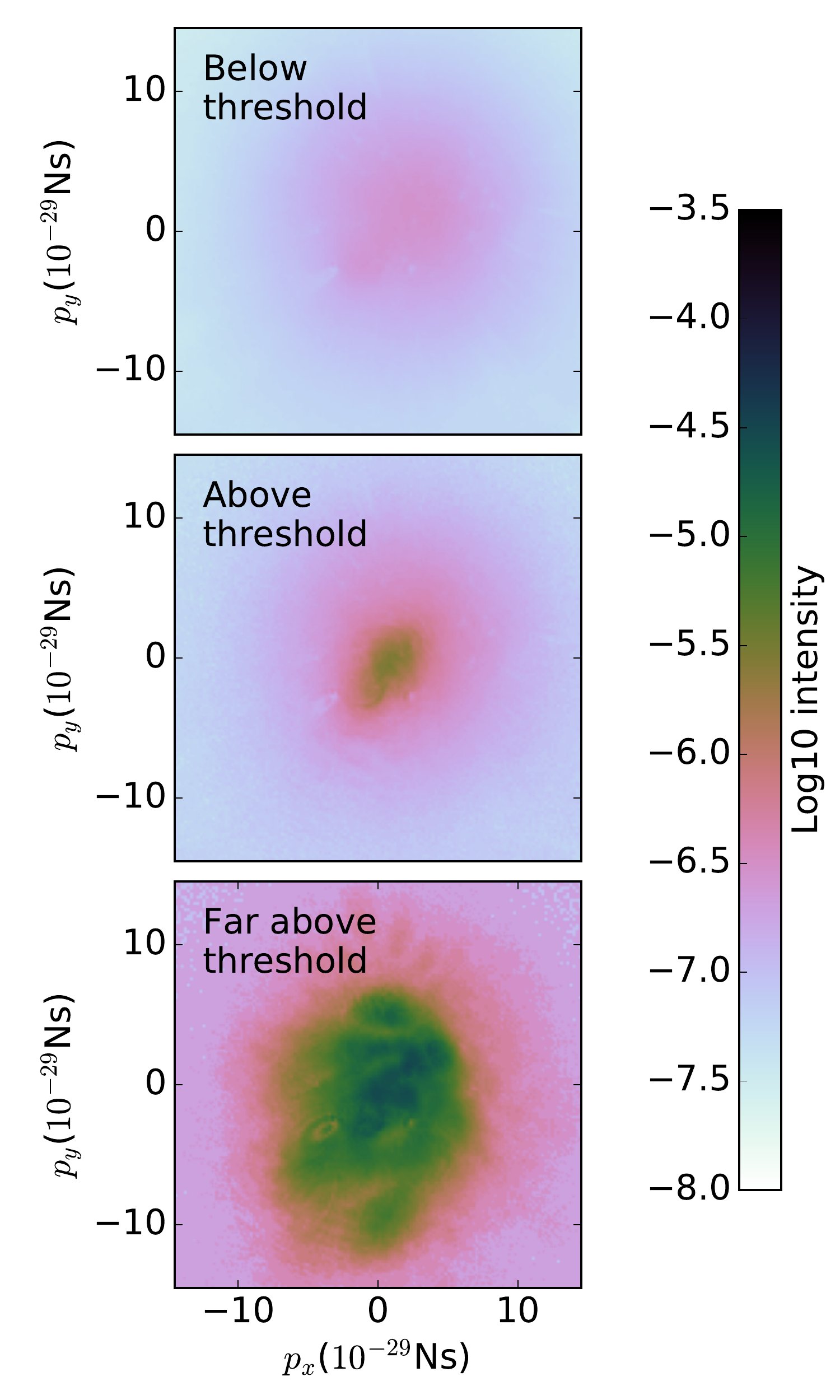}
	\caption{Momentum-space ${\bf p}=(p_x, p_y)$ images, $n({\bf p})$. As pump power is increased (top to bottom), to the thermal distribution (top panel) is added a momentum-narrow condensate (middle), followed by a broadened multi-mode condensate (bottom panel). The calibration is accurate for $|{\bf p}|<4\times10^{-29}$~Ns.}
	\label{fig:momentum space images}
\end{figure}

Using the diagnostic camera, we obtain momentum-space images $n({\bf p})$, as shown in \figref{fig:momentum space images}. A broad thermal cloud is seen. Above threshold, a condensate peak appears on top of the thermal cloud. At very high pump powers, multimode condensation occurs, which leads to a distorted, broadened condensate peak. The same sorts of peaks are observed in position-space imaging, also broadening when multimode condensation is seen. Since position-space and momentum-space are mutually Fourier transforms, we conclude that the multiple modes are incoherent, in agreement with Ref.~\cite{Marelic16}.

\begin{figure}[htb]
	\centering
	\includegraphics[width=0.9\columnwidth]{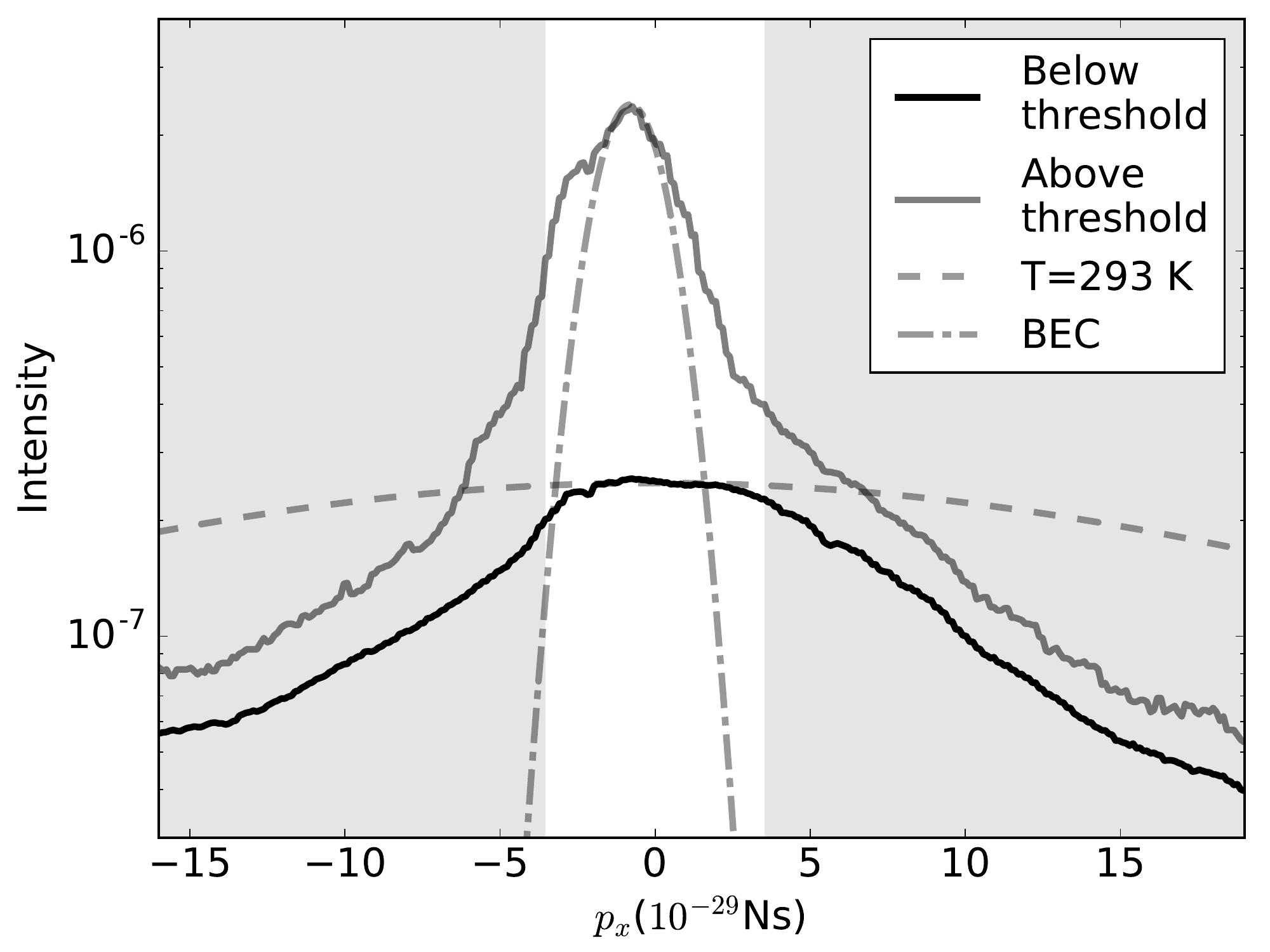}
	\caption{Cuts through the same data as \figref{fig:momentum space images}, top two panels. The greyed out region is known to suffer from pincushion distortion, rendering the calibration inaccurate. The dashed line is a thermal distribution at room temperature. The dot-dashed line is the expected momentum distribution for the non-interacting Bose-Einstein condensate.}
	\label{fig:momentum space cuts}
\end{figure}

The width of the condensate peak is approximately equal to the characteristic momentum width of the harmonic oscillator $p_{ho} = \sqrt{\hbar \Omega m_{ph}}$, since photon-photon interactions are negligible. For the experiment of \figref{fig:momentum space images}, $p_{ho} = 1.1\times10^{-29}$~Ns. The typical thermal momentum scale for temperature $T$ is $\sqrt{m_{ph} k_B T} = 1.8\times10^{-28}$~Ns. In \figref{fig:momentum space cuts} we see cuts through the momentum space images just below and above condensation threshold (the same data as the top two panels of \figref{fig:momentum space images}). The greyed out sections show significant distortion, so the calibration therein of vertical momentum $p_y$ is not trustworthy. Gaussian curves of widths appropriate to the thermal (dashed line) and the condensate (dash-dotted line) momentum size are also shown. In the calibrated region the width of the condensate is compatible with the harmonic oscillator momentum width and the below-threshold momentum width is compatible with the typical thermal momentum scale.

Repulsive interactions would increase the position-space size of the condensate, which decreases the momentum-space size. We have seen no such effect, which confirms our bound of $\tilde{g} \lesssim 10^{-3}$ based on the dispersion relation.

\section{Position space}

We now turn to position-space resolved spectra. Where the cavity length locally varies, the energy of light which matches the mirror boundary conditions also varies locally. Thus, mirror shape is mapped to potential energy landscape, $V({\bf r})$.  Similar spectra have been demonstrated for condensates of exciton-polaritons when a confining potential is applied~\cite{Balili07}.

\subsection{Calibration of position}

One of the two cavity mirrors is planar, the other spherical with a radius of curvature $R=0.25$~m. The light penetrates the dielectric mirrors, which have a different refractive index to the dye-solvent mixture, which means that the optical path length is not exactly the same as the true cavity length. We write the cavity length as a function of longitudinal mode number $q$ so that 
\begin{align}
  L_0(q) = \frac{\lambda^*q}{2} - L_{off}
    \label{eqn:Lq}
\end{align}
where $L_{off}$ is the distance the light penetrates the mirrors and $\lambda^* = \lambda_0 / n$ is the wavelength of light in the intracavity medium. In principle $L_{off}$ is also a function of wavelength, but in practice it makes no difference if we take it to be a constant here.

\begin{figure}[htb]
  \centering
  \includegraphics[width=0.8\columnwidth]{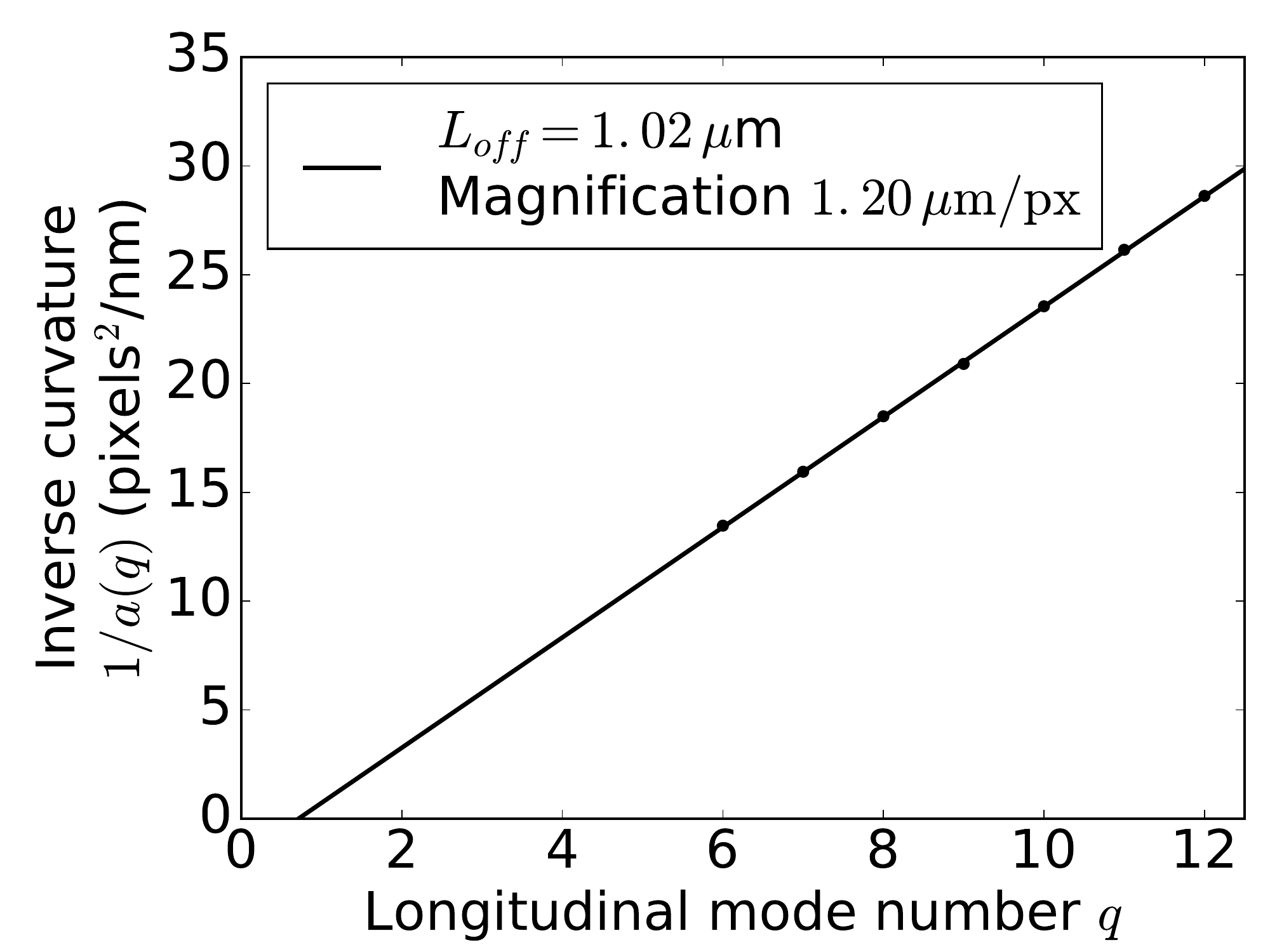}
  \includegraphics[width=0.8\columnwidth]{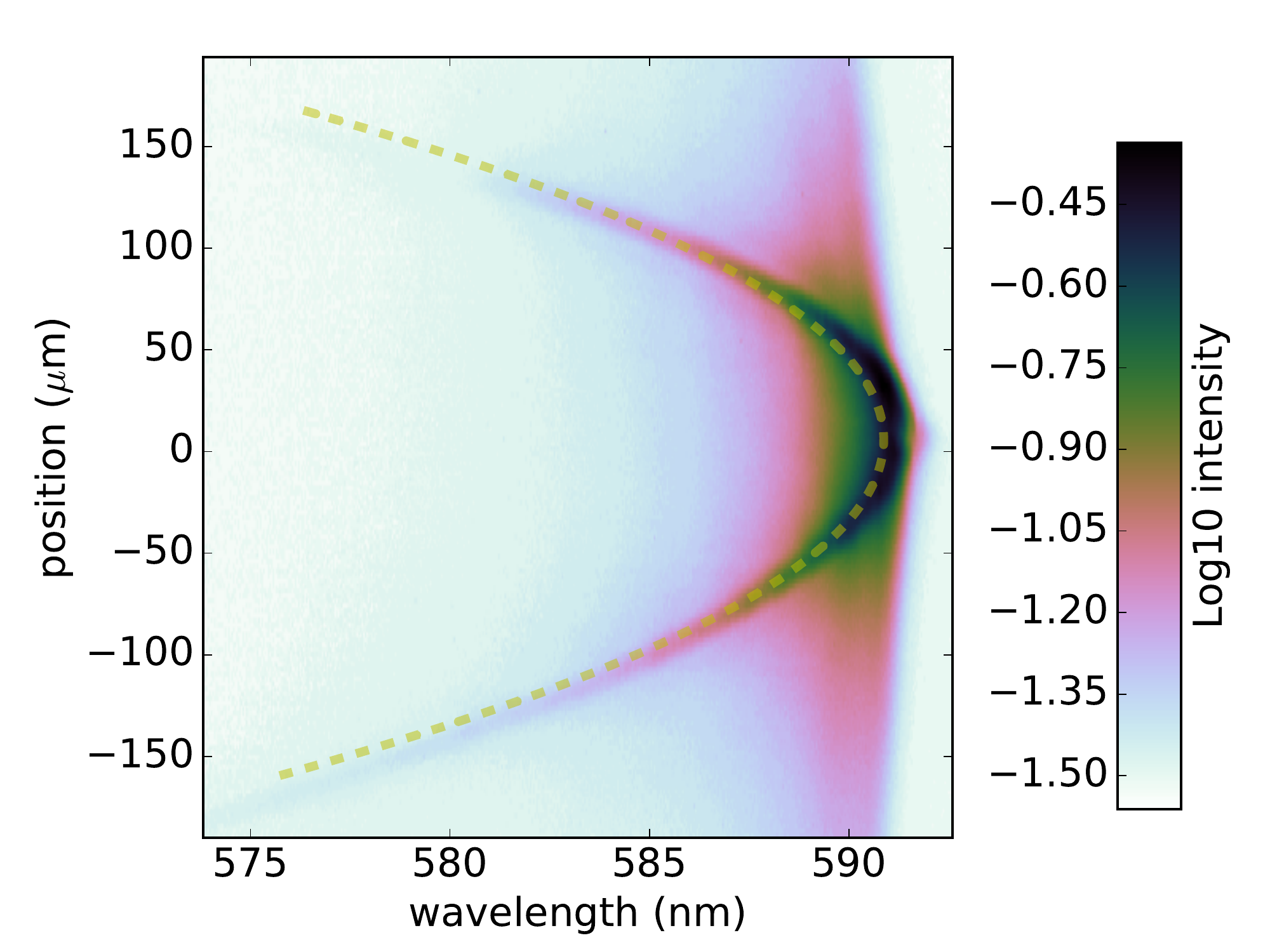}
  \caption{Top: The measured parabolic curvature $a$ of the position dependence of the confining potential for photons depends on the longitudinal mode number (dots). Data were all taken with cutoff $\lambda_0 = 591\pm 0.7$~nm. A fit to \eqnref{eqn:Lq} (solid line) allows us to calibrate the magnification in the position-resolved spectra. Part of the light penetrates the cavity mirrors, to an optical depth equivalent to $L_{off}$, also obtained from the fit. Bottom: an example calibrated position-resolved spectrum $n({\bf r},E)$ with the fitted parabolic potential.}
  \label{fig:position calibration}
\end{figure}

We can use \eqnref{eqn:Lq} to calibrate the position-space magnification of the position-resolved spectra. We expect the potential energy landscape to appear as a parabolic cutoff energy as a function of position, as shown in \eqnref{eqn:frE}. The parabolic coefficient in object space $V(r) = a r^2$ is given by:
\begin{align}
  \frac{1}{a(q)}  = {R}\left(\frac{\lambda_0}{h c}\right)  
      \left[ {\lambda^*q} - 2 L_{off}\right]
	\label{eqn:parabola}
\end{align}
The energy calibration of the spectrometer is performed as for the momentum-resolved spectroscopy. In \figref{fig:position calibration} (top) we show the measured inverse of the parabolic curvature $a$ in nm~per~(camera~pixel)$^2$ for various longitudinal mode numbers. From the fit to \eqnref{eqn:parabola} both the mirror penetration and magnification of the system are determined. An example position-resolved spectrum with magnification calibrated is shown in \figref{fig:position calibration} (bottom). The fitted parabola is overlaid (dashed line).

\subsection{Position-resolved spectra $n({\bf r},E)$ and images $n({\bf r})$}

\begin{figure*}[htb]
	\centering
	\includegraphics[width=1.35\columnwidth]{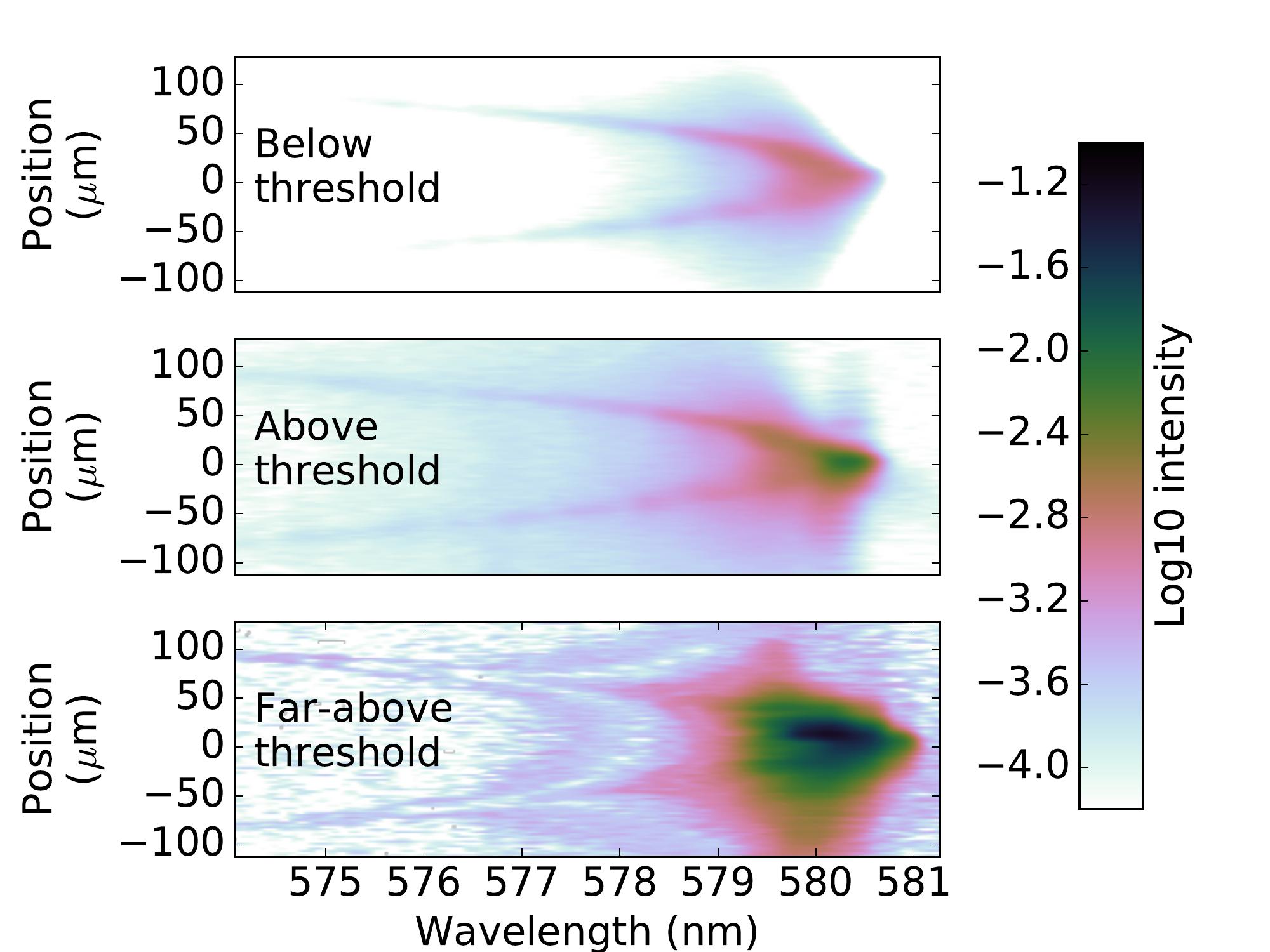}
	\includegraphics[width=0.6\columnwidth]{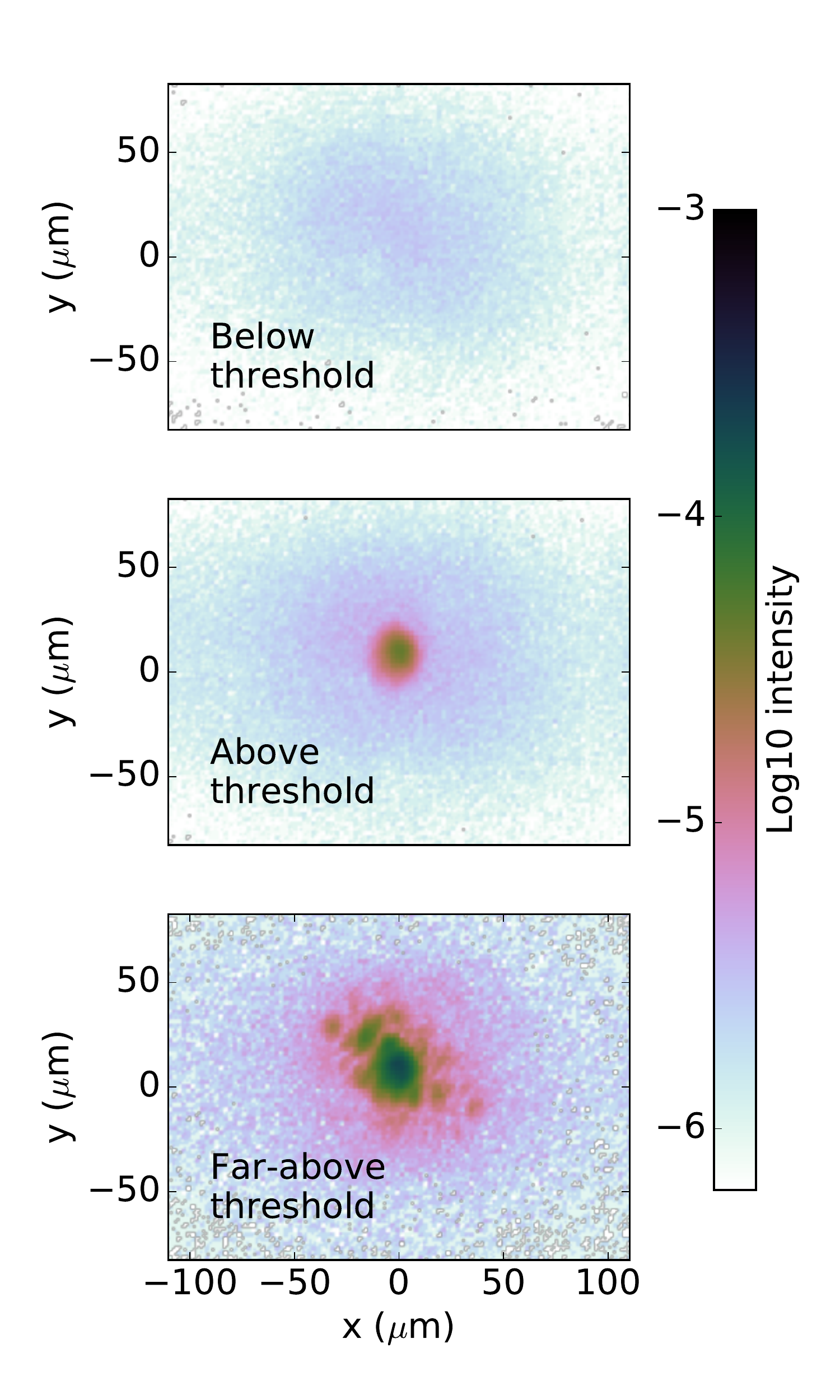}
	\caption{
	Position-resolved spectra $n({\bf r},E)$ (left side) and images $n({\bf r})$ with ${\bf r}=(x,y)$ (right side) of photoluminescence below threshold (top) and BEC just above (middle) and far above threshold (bottom). The parabolic curve of the harmonic potential energy is clearly visible. Condensation occurs at the centre, where potential energy is lowest. At higher pump powers, multimode condensation is observed.
	}
	\label{fig:condensate position resolved}
\end{figure*}

We show in \figref{fig:condensate position resolved} (left) typical position-resolved spectra $n({\bf r},E)$ of a our dye-microcavity photoluminescence far below threshold, just above threshold for single-mode condensation and far above threshold for multimode condensation. On the right are position-space images of the same conditions.

The photoluminescence is most intense near the potential energy, whereas \eqnref{eqn:frE} predicts uniform intensity for each energy, for all energies higher than the minimum set by the potential energy. The explanation is that only photoluminescence with low kinetic energy, i.e. small emission angle, is admitted by our optics. The slit in the telescope, positioned at a real-space image, acts in part as an anisotropic momentum-space filter. A side effect is that the position-resolution is no better than about 15~\micron\ in the resolved spectrometer.


\section{Discussion}

\subsection{Limitations of the optical setup}

The experiment was conceived in an attempt to measure the Bogoliubov dispersion relation which gives information about the photon-photon interactions. In Ref.~\cite{Nyman14} it was shown that wavelength energy resolution 0.04~nm and momentum resolution $1.4\times10^{-30}$~N\,s would be sufficient to measure a dimensionless interaction parameter $\tilde{g} = 10^{-5}$, which is in the expected range~\cite{VanDerWurff14}. The measured wavelength resolution depends on the mechanical stability of the cavity, which ranges from 0.05~nm (for $q$ so small that the mirrors touch) to 0.5~nm (in more normal operation). The minimum position and momentum resolution we have achieved here are 15~\micron\ and $4\times10^{-30}$~N\,s respectively. For comparison, the smallest cavity mode covers 6~\micron\ and $1.1\times10^{-29}$~N\,s: we resolve well in momentum but not position. The wavelength resolution is sufficient to resolve interactions at the $10^{-5}$ level, but not so momentum. The multimode condensation phenomenon also makes interpretation in terms of interactions tricky.

The resolution limits in position are set mainly by spherical aberrations in the telescope. The achromatic doublets used give distortions for angles greater than about 35~mrad~\cite{BornWolf}, equivalent to $4\times10^{-29}$~N\,s, exactly where our calibration is seen to become invalid. Resolution could be improved by replacing the achromatic doublets with lenses which correct more aberrations or by removing the telescope altogether. The latter option would require a slit in the back focal plane of the objective, precluding simultaneous momentum- or position-resolved spectroscopy and imaging as presented in \figref{fig:condensate position resolved}. In addition, momentum-space imaging optics could be moved closer, occupying a large numerical aperture, improving the momentum-space resolution.

The momentum resolution is in practice limited by the inverse of the size of the lowest cavity mode. To improve effective momentum resolution we would therefore use larger radius of curvature mirrors. More gently curved mirrors also cause there to be more photons at threshold, enhancing the effects of interactions, at the cost of decreasing the energy range over which interaction effects occur. The energy resolution can be improved to compensate by using a more finely spaced grating than the present 1800 lines/mm.

\subsection{Conclusions}

We have observed various projections of the phase-space distribution of dye-microcavity thermalised and condensed photons: position- and momentum-resolved, with and without energy resolution. The momentum-resolved spectrum gives the dispersion relation, which results in upper bounds for both the photon-photon and photon-molecule interaction strengths. Both are very weak, indicating that the light is acting as free photons. Reasonable improvements to the experimental setup might well make it possible to directly measure the photon-photon interaction strength, which is still not well known for dye-microcavity photons. Since we find that the effective temperature is not a phase-space local quantity, it is clear that ergodicity is retained even when the temperature is not room temperature because of incomplete thermalisation.

\section{Acknowledgements}

We thank the UK Engineering and Physical Sciences Research Council for supporting this work through fellowship EP/J017027/1 and the Controlled Quantum Dynamics CDT EP/L016524/1. The data underlying this article, as well as the scripts used for analysis, are available at \url{http://dx.doi.org/10.5281/zenodo.60901}.


\bibliographystyle{prsty}
\bibliography{photon_bec_refs}

\clearpage

\end{document}